\documentclass[12pt]{article}

\usepackage{amssymb}
\usepackage{color}
\usepackage{epsfig,amssymb,amsfonts,amsmath,graphicx,dsfont,cite,xfrac}
\usepackage{authblk}
\usepackage{subfigure}
\definecolor{mygray}{gray}{0.5}

\usepackage{cite}
\usepackage[colorlinks=true,linkcolor=blue,citecolor=red]{hyperref}
\newcommand{\be}{\begin{equation}}
\newcommand{\ee}{\end{equation}}
\newcommand{\bea}{\begin{eqnarray}}
\newcommand{\eea}{\end{eqnarray}}

\title{Completeness and Nonclassicality of Coherent States for Generalized Oscillator Algebras}

\author[1]{Kevin Zelaya}
\author[1]{Oscar Rosas-Ortiz\thanks{Corresponding author. E-mail: orosas@fis.cinvestav.mx (O. Rosas-Ortiz).}}
\author[1]{Zurika Blanco-Garcia}
\author[2]{Sara Cruz~y~Cruz}

\affil[1]{\footnotesize Physics Department, Cinvestav, AP 14-740, 07000
M\'exico City, Mexico}

\affil[2]{\footnotesize Instituto Polit\'ecnico Nacional, UPIITA, Av. I.P.N. 2580, Col. La Laguna Ticom\'an, C.P. 07360, M\'exico City, Mexico}

\date{}
\begin{document}

\maketitle

\begin{abstract}

The purposes of this work are (1) to show that the appropriate generalizations of the oscillator algebra permit the construction of a wide set of nonlinear coherent states in unified form; and (2) to clarify the likely contradiction between the nonclassical properties of such nonlinear coherent states and the possibility of finding a classical analog for them since they are $P$-represented by a delta function. In (1) we prove that a class of nonlinear coherent states can be constructed to satisfy a closure relation that is expressed uniquely in terms of the Meijer $G$-function. This property automatically defines the delta distribution as the $P$-representation of such states. Then, in principle, there must be a classical analog for them. Among other examples, we construct a family of nonlinear coherent states for a representation of the $su(1,1)$ Lie algebra that is realized as a deformation of the oscillator algebra. In (2), we use a beam splitter to show that the nonlinear coherent states exhibit properties like anti-bunching that prohibit a classical description for them. We also show that these states lack second order coherence. That is, although the $P$-representation of the nonlinear coherent states is a delta function, they are not full coherent. Therefore, the systems associated with the generalized oscillator algebras cannot be considered `classical' in the context of the quantum theory of optical coherence.

\end{abstract}

%%%%%%%%%%%%%%%%%%%%%%%%%%%%%%%%%%%%%%%%%%%%%%%

%---------------------------------------> Section
\section{Introduction}

The nonclassical properties of light have received a great deal of  attention in recent years, mainly in connection with quantum optics \cite{Dod03}, quantum information \cite{Bou00},  and the principles of quantum mechanics \cite{Sil08}. Among other nonclassical profiles, the related states can be generated by  nonlinear processes to have sub-Poissonian statistics \cite{Man79} or to exhibit squeezing (reduction of the variance) in one quadrature \cite{Wal83}. Some deformations of the algebra generated by the conventional boson operators have been proposed to represent photons with `unusual properties' \cite{Bie89,Mac89}, which have found applications in the photon counting statistics, squeezing and signal-to-quantum noise ratio \cite{Sol94}. The immediate generalizations \cite{Mat96,Man97} motivated the development of the subject as an important branch of quantum optics \cite{Dod03}. The deformations of the boson algebra include supersymmetric structures \cite{Mie04,Aoy01a,Aoy01b,And12} for which the so-called polynomial Heisenberg algebras are quite natural \cite{Mie84,Fer95,Ros96,Fer07}. Recently, some non-Hermitian models have been shown to obey the distortions of the boson algebra that arise in the conventional supersymmetric approaches \cite{Ros15,Ros17,Zel17}. In all cases, the deformed oscillator algebras have been used to construct the corresponding generalized (also called nonlinear) coherent states. Most of these states exhibit nonclassical properties that distinguish them from the coherent states of the conventional boson algebra.

In this work we propose a modification of the conventional boson (also called oscillator) algebra that permits the recovering of the majority of the already studied deformed algebras as particular cases. The nonlinear coherent states of these generalized oscillator algebras can be written in the same mathematical form, which facilitates their study. We shall address the discussion to polynomial-like algebras because, as we are going to prove, the closure relation satisfied by the corresponding nonlinear coherent states is expressed uniquely in terms of the Meijer $G$-function. Accordingly, the $P$-representation \cite{Gla63,Sud63} of all the nonlinear coherent states we deal with is as singular as the $\delta$-distribution. Then, following \cite{Gla07,Kla68}, the fields represented by such states would have a classical analog. However, these states have nonclassical properties that can be exhibited either with the help of a beam splitter \cite{Kim02} (see also \cite{Wan02}) or by showing that their statistics is sub-Poissonian \cite{Man79}. We face this likely contradiction by interpreting the action of a beam splitter on a given photon state as the equivalent of a double-slit interference experiment in the single-photon regime \cite{Gra86,Rue13} (see also \cite{Han56}). Then, we use this equivalence to show that the fields represented by the nonlinear coherent states exhibit anti-bunching (the probability of two or more photons arriving together at the same point is zero), so they cannot be modeled in classical form. The conclusion is that the nonlinear coherent states are not full coherent although their $P$-representation is a delta function. Then, they do not satisfy the notion of classicalness introduced by Glauber in his quantum theory of optical coherence \cite{Gla07}.

The paper is organized as follows. In Sec.~\ref{oscillators} we generalize the oscillator algebras and give some immediate examples. Sec.~\ref{S_nonlinear} deals with the construction of the nonlinear coherent states. We show that the completeness of these states is always possible for the polynomial-like algebras in terms of the Meijer $G$-function. Some immediate examples are also discussed. In Sec.~\ref{hilbert} we show that the $P$-representation of the nonlinear coherent states is given by the $\delta$-distribution while the photon states are represented by the derivatives of $\delta$. In Sec.~\ref{fundamento} we analyze the likely contradiction between the completeness and the nonclassicality of the nonlinear coherent states. Some final remarks are given in Sec.~\ref{conclu}. We have added an appendix with some important mathematical expressions that are not required for reading the paper but are necessary to follow the calculations. 
 
%---------------------------------------> Section
\section{Generalized oscillators}
\label{oscillators}

The conventional boson ladder operators $\hat a$ and $\hat a^{\dagger}$ satisfy the algebraic relation $[\hat a, \hat a^{\dagger}]=1$. If the number operator $\hat n = \hat a^{\dagger} \hat a$ is considered, then $[ \hat n, \hat a]=-\hat a$ and $[\hat n, \hat a^{\dagger}]=\hat a^{\dagger}$. The latter expressions define the oscillator (or boson) algebra and show that the action of $\hat a^{\dagger}$ on any number eigenvector $\vert n \rangle$, $n=0,1,2,\ldots$, produces a new eigenvector of $\hat n$ with eigenvalue $n+1$. Similarly, $\hat a \vert n +1 \rangle$ is proportional to the number eigenvector $\vert n \rangle$, while $\hat a \vert 0 \rangle =0$. In the sequel we shall modify the oscillator algebra by preserving the number operator $\hat n$ but changing the ladder operators, now written $\hat a_E$ and $\hat a_E^{\dagger}$, as the set of generators. We say that any system obeying the new algebra is a generalized oscillator.

%---------------------------------------> Subsection
\subsection{Deformed oscillator algebras}

To define the generalized oscillators let us introduce the pair of ladder operators 
\be
\hat a_E \vert n \rangle = \sqrt{E(n)} \vert n-1 \rangle, \quad \hat a^{\dagger}_E \vert n \rangle = \sqrt{E(n+1)} \vert n+1 \rangle, \quad n=0,1,2, \ldots,
\label{alg1}
\ee
where $E$ is a nonnegative function,  $\hat a_E^{\dagger}$ is the Hermitian conjugate of $\hat a_E$, and $\vert n \rangle$ is eigenvector of the number operator with eigenvalue $n$. The following algebraic relations can be proven
\be
[\hat n, \hat a_E]= -\hat a_E, \quad [\hat n, \hat a^{\dagger}_E]= \hat a^{\dagger}_E.
\label{alg2}
\ee
The product $\hat a_E^{\dagger}  \hat a_E$ preserves the number of quanta since it is equal to the function $E(\hat n)$. Equivalently, $\hat a_E \hat a^{\dagger}_E=E(\hat n +1)$. Then
\be
[\hat a_E, \hat a^{\dagger}_E] = E(\hat n+1) - E(\hat n).
\label{alg3}
\ee
As the vacuum state $\vert 0 \rangle$ does not contain quanta we shall assume $E(0)=0$ in order to have $\hat a_E \vert 0 \rangle =0$. The number eigenvectors
\[
\vert n \rangle = \frac{ ( a^{\dagger} )^n }{\sqrt{n!} } \vert 0 \rangle, \quad n=0,1,2, \ldots,
\]
are now expressed as
\be
\vert n \rangle = \frac{ ( a^{\dagger}_{E \,} )^n }{ \sqrt{ E(n)! }} \,  \vert 0 \rangle, \quad E(n)!= E(1) E(2) \cdots E(n), \quad E(0)! \equiv1, \quad n = 0,1, \ldots,
\label{state1}
\ee
and will be used as the orthonormal basis of the Hilbert space ${\cal H}$. The latter consists of all vectors
\be
\vert \psi \rangle = \sum_{n=0}^{+\infty} \psi_n \vert n \rangle, \quad \psi_n := \langle n \vert \psi \rangle \in \mathbb C,
\label{state2}
\ee
such that  
\be
\langle \psi \vert \psi \rangle = \sum_{n=0}^{+\infty} \vert \psi_n \vert^2 < \infty.
\label{state3}
\ee

%---------------------------------------> Subsection
\subsection{Examples}

The following list of examples do not exhaust all the possible generalizations of the oscillator algebras that can be performed with the rules defined in Eqs.~(\ref{alg1})--(\ref{alg3}). We explicitly mention such cases because either they are connected with approaches already reported, or they give rise to very important results in mathematical physics or they find applications in quantum optics. 

%---------------------------------------> Subsubsection
\subsubsection{$f$-oscillators}
\label{fosc}

An important class of generalized harmonic oscillators has been already introduced and is nowadays known as  the set of $f$-oscillators \cite{Man97}. These oscillators are recovered here by making $E(\hat n) = \hat n f^2 (\hat n)$ in (\ref{alg1})--(\ref{alg3}), with $f$ a properly chosen real-valued function. In this case the ladder operators $\hat a_E$ and $\hat a_E^{\dagger}$ are factorized as 
\be
\hat a_E = \hat a f(\hat n) = f(\hat n +1) \hat a, \quad \hat a_E^{\dagger} = f(\hat n) \hat a^{\dagger}  = \hat a^{\dagger} f(\hat n +1).
\label{foperators}
\ee
The $f$-oscillators have been associated with the center-of-mass motion of a trapped and bichromatically laser-driven ion \cite{Mat96}, and are also related to a `frequency blue shift' in high intensity photon beams \cite{Man97}.

%---------------------------------------> Subsubsection
\subsubsection{$q$-deformed oscillators}
\label{qosc}

It has been shown \cite{Man97} that the special choice of the $f$-function
\be
f(\hat n)= \sqrt{ \frac{ \sinh (\lambda \hat n)}{ \hat n \sinh \lambda}}, \quad \lambda = \ln q, \quad q \in \mathbb R,
\ee
reduces the commutator (\ref{alg3}) of the $f$-operators (\ref{foperators}) to the rule of the $q$-deformed oscillators obeyed by the ``physics'' bosons \cite{Bie89,Mac89}
\be
\hat a_E \hat a_E^{\dagger} - q \hat a_E^{\dagger} \hat a_E = q^{-\hat n}.
\ee
The above approach has important applications in quantum optics as regards the photon counting statistics, squeezing and signal-to-quantum noise ratio \cite{Sol94}.

%---------------------------------------> Subsubsection
\subsubsection{Polynomial-like oscillators}
\label{polosc}

Other relevant class of oscillators is obtained by assuming that $E( \hat n)$ is a real polynomial of degree $\ell$,
\be
E( \hat n)= \prod_{p=1}^{\ell} (\alpha_p \hat n + \beta_p) = \gamma_{\ell} \prod_{p=1}^{\ell} ( \hat n+ \delta_p), \quad \gamma_{\ell} = \prod_{p=1}^{\ell} \alpha_p, \quad \delta_p = \frac{\beta_p}{\alpha_p}.
\label{epol1}
\ee
The expression 
\be
E(n)! = \gamma_{\ell}^n \prod_{p=1}^{\ell}  \frac{ \Gamma(n+1 + \delta_p) }{\Gamma(1+\delta_p)}
\label{epol2}
\ee
is easily achieved and will be useful in the sequel. In this case the commutator relations (\ref{alg2})--(\ref{alg3}) define a {\em polynomial Heisenberg algebra} of degree $\ell -1$. The polynomial algebras are quite natural in the higher order supersymmetric approaches \cite{Mie04,Aoy01a,Aoy01b,And12}, and are usually connected with nonlinearities that arise because the differential order of the operators that intertwine the susy partner Hamiltonians is greater than one \cite{And12}. The examples discussed in Secs.~\ref{S_su11}, \ref{secsusy} and \ref{secdist}, below, are special cases of polynomial algebras.

$\bullet$ The simplest example is obtained for $\ell=1$. The commutator (\ref{alg3}) is the $0$--degree polynomial $E(\hat n+1) - E(\hat n)= \alpha_1$. The condition $E(0)=0$ implies $\beta_1=0$, by necessity. Then, for $\alpha_1=1$ one gets $E(\hat n)=id$, with $id \equiv \mathbb I$ the identity operator in ${\cal H}$. In this case $\hat a_{id} =\hat a$ and $\hat a_{id}^{\dagger} = \hat a^{\dagger}$, so that the commutation rules (\ref{alg2})--(\ref{alg3}) define the conventional algebra of the oscillator.

%---------------------------------------> Subsubsection
\subsubsection{$su(1,1)$-oscillators}
\label{S_su11}

If $E$ is a quadratic polynomial ($\ell=2$), the commutator (\ref{alg3}) gives rise to the first order degree polynomial
\be
E(\hat n+1)-E(\hat n)= 2 \alpha_1 \alpha_2 \hat n + \alpha_1 \beta_2 + \alpha_2 \beta_1 + \alpha_1 \alpha_2.
\ee
A striking example occurs for $\beta_2=0$ since the related function\footnote{Similar expressions are obtained for $\beta_1=0$ and arbitrary $\beta_2$.} $E(\hat n)= \alpha_1 \alpha_2 \hat n^2 + \alpha_2 \beta_1 \hat n$ leads to the $su(1,1)$ Lie algebra
\be
[K_0, K_{\pm}]= \pm K_{\pm}, \quad [K_-, K_+]=2K_0.
\label{su11}
\ee
Indeed, by making $E(\hat n)= \hat n f^2(\hat n)$ with $f(\hat n) =\sqrt{\alpha_2 (\alpha_1 \hat n+ \beta_1)}$, the identification 
\be
\begin{array}{c}
K_-= \hat a_E = \hat a \sqrt{ \alpha_2 (\alpha_1 \hat n+ \beta_1) }, \quad K_+=\hat a_E^{\dagger}= \sqrt{ \alpha_2 (\alpha_1 \hat n+ \beta_1) } \, \hat a^{\dagger},\\ [2ex] 
K_0 = \alpha_1 \alpha_2 \hat n +\frac12 (\alpha_2 \beta_1 + \alpha_1 \alpha_2),
\end{array}
\ee
gives the algebra (\ref{su11}) from the relations (\ref{alg2})--(\ref{alg3}).

$\bullet$ If additionally to $\beta_2=0$ we take $\alpha_1 = \alpha_2 = \beta_1 = 1$, then $E(\hat n)=\hat n (\hat n+1)$. The generators of the $su(1,1)$ Lie algebra are in this case $K_0= \hat n +1$, $K_- = \hat a \sqrt{ \hat n +1}$ and $K_+ = K_-^{\dagger}$. The operators $K_{\pm}$ have been already used to represent the atom-photon coupling of the Jaynes-Cummings model \cite{Jay63} for  intensity dependent interactions \cite{Buc81,Suk81}.

%---------------------------------------> Subsubsection
\subsubsection{SUSY-like oscillators}
\label{secsusy}

A special case of the commutator (\ref{alg3}) is obtained if the $E$-function is a cubic polynomial ($\ell=3$) such that $E(1)=E(0)=0$. The latter condition means that the vacuum $\vert 0 \rangle$ is annihilated by both ladder operators, $\hat a_E$ and $\hat a_E^{\dagger}$, while the $1$-photon state $\vert 1 \rangle$ is annihilated by $\hat a_E$. That is
\be
\hat a_E \vert 1 \rangle = \hat a_E \vert 0 \rangle= \hat a_E^{\dagger} \vert 0 \rangle=0.
\label{susy1}
\ee
Then, the action of the ladder operators on the subspace $\{ \vert n +2 \rangle, n=0,1,2,\ldots \}$ is a modification of the rule (\ref{alg1}), namely
\be
\hat a_E \vert n +2 \rangle = \sqrt{E(n+2)} \vert n+1 \rangle, \quad \hat a^{\dagger}_E \vert n +1 \rangle = \sqrt{E(n+2)} \vert n+2 \rangle, \quad n \geq 0.
\label{susy2}
\ee
Assuming that the solutions of (\ref{susy1}) are given, the number eigenvectors $\vert n +2 \rangle$ are now generated from the 1-photon state 
\be
\vert n+2 \rangle = \frac{ \hat a_E^{\dagger (n+1)} }{ \sqrt{ E(n+2)!} } \vert 1 \rangle, \quad n\geq 0.
\label{susy3}
\ee
The above construction is associated with the eigenstates of a series of Hermitian Hamiltonians that share their spectrum with the conventional harmonic oscillator, which is shifted in one unit of energy. That is, the energies are given by $E_n =n-\sfrac12$ \cite{Mie84}. Such Hamiltonians are supersymmetric (SUSY) partners of the conventional oscillator for which the supersymmetry is {\em unbroken} \cite{Mie04}. A more general treatment includes non-Hermitian Hamiltonians whose eigenvalues are given by $E_{n+1}=n +\sfrac12$, and $E_0 =\epsilon < \sfrac12$ \cite{Ros15,Ros17}. That is, the spectrum of such non-Hermitian Hamiltonians includes all the oscillator energies $n+\sfrac12$ plus an additional real eigenvalue $\epsilon$ which is located below the ground state energy of the conventional oscillator. The $E$-function for this case can be derived from (\ref{epol1}) with $\alpha_1 =\alpha_2 =\alpha_3=\beta_1=1$, $\beta_2=  \frac12- \epsilon$, and $\beta_3 = \frac32-\epsilon$. Thus,
\be
E(\hat n+2)= (\hat n+1) \left( \hat n+\frac12 -\epsilon \right) \left( \hat n+ \frac32 -\epsilon \right), \quad n\geq 0.
\label{esusy}
\ee
The non-Hermitian Hamiltonians associated to (\ref{esusy}) are constructed with a complex-valued potential \cite{Ros15}, so that their study requires a bi-orthogonal structure for the space of states \cite{Ros17,Zel17}. In the appropriate limit, the imaginary part of the potential is cancelled and the model becomes Hermitian, although the generalized algebra defined by (\ref{esusy}) is preserved. Then, in such limit, the algebraic structure of the oscillators reported in \cite{Mie84} is recovered by making $\epsilon =-\sfrac12$ in (\ref{esusy}).

Notice that the algebras described above depend on the ground energy of the system. That is, systems with different ground energies $\epsilon$ will be regulated by different algebras. 

%---------------------------------------> Subsubsection
\subsubsection{Distorted SUSY-like oscillators}
\label{secdist}

Another $E$-function that satisfies the supersymmetric relations (\ref{susy1})--(\ref{susy3}) is defined as $E(\hat n+2) = w + \hat n$, with $w$ a nonnegative parameter and $E(1)=E(0)=0$. In this case the commutator (\ref{alg3}) gives
\be
E(\hat n+1)-E(\hat n) = \left\{
\begin{array}{rl}
0 & n=0\\[1ex]
w & n=1 \\[1ex]
1 & n\geq 2
\end{array}
\right.
\label{distorted1}
\ee
Thus, the operators $\hat a_E$ and $\hat a_E^{\dagger}$ are the generators of an algebra that imitates the Heisenberg one. For if we concentrate on the subspace spanned by $\{ \vert n \rangle, n\geq 2 \}$ only, then the commutator (\ref{distorted1}) is completely equivalent to the oscillator one. The same occurs in the subspace spanned by $\vert 1 \rangle$, up to the constant $w \geq 0$. For this reason, the algebra defined by (\ref{distorted1}) is referred to as  {\em distorted Heisenberg algebra}, and $w$ is called the {\em distortion parameter} \cite{Fer95,Ros96}. 

We would like to remark that, contrary to what happens with the algebra of the previous section, the distorted Heisenberg algebra does not depend on the ground energy of the system. Besides, the results associated to (\ref{distorted1}) are also easily extended to non-Hermitian Hamiltonians by using the bi-orthogonal approach developed in \cite{Ros17,Zel17}. 

Other versions of polynomial algebras can be obtained from either the ${\cal N}$--fold or the nonlinear supersymmetric models discussed in \cite{Aoy01a,Aoy01b} and \cite{And12} respectively.

%---------------------------------------> Section
\section{Nonlinear coherent states}
\label{S_nonlinear}

Up to a normalization constant, the solutions to the eigenvalue equation $\hat a_E \vert z_E \rangle = z \vert z_E \rangle$, with $z \in \mathbb C$, can be written in the form
\be
\vert z_E \rangle =  \sum_{n=0}^{+\infty} \frac{z^n}{\sqrt{ E(n)! }} \vert n \rangle.
\label{cs1}
\ee
The vector $\vert z_E \rangle$ belongs to ${\cal H}$ only if the series (\ref{cs1}) is norm convergent. Thus, if $\vert z_E \rangle_N = {\cal N}_E ( \vert z\vert)  \vert z_E \rangle$ denotes the normalized solution ${}_N \langle z_E \vert z_E \rangle_N= \vert {\cal N}_E( \vert z \vert) \vert^2 \langle z_E \vert z_E \rangle= 1$, then the expression
\be
{\cal N}_E ( \vert z \vert)  = \left[ \sum_{n=0}^{+\infty} \frac{ \vert z \vert^{2n}}{E(n)!}
\right]^{-1/2}
\label{cs2}
\ee
must be finite. Clearly, not any $E$ and $\vert z \vert$ are allowed. In the following we assume that  ${\cal N}_E ( \vert z \vert)$ is finite for $z \in {\cal S} \subseteq \mathbb C$, with  ${\cal S}$ defined whenever $E$ has been provided, and say that $\vert z_E \rangle_N$ is a generalized (nonlinear) coherent state. Concrete realizations will be shown in the examples. 

The probability of having $n$ photons associated to the nonlinear coherent state $\vert z_E \rangle_N$ is given by ${\cal P}_E (n, \vert z \vert)= \vert \langle n \vert z_E \rangle_N \vert^2= \frac{  {\cal N}_E^2(\vert z \vert) \vert z \vert^{2n} }{E(n)! }$,
and the average photon number is
\be
\langle \hat n \rangle_{z_E} = {\cal N}_E^2(\vert z \vert) \sum_{n=0}^{+\infty} \frac{ \vert z \vert^{2(n+1)}}{E(n+1)! } (n+1).
\label{expec1}
\ee
Following \cite{Jac10} we may introduce the $E$-exponential function (see details in Appendix~\ref{ApA})
\be
e^x_E = \sum_{n=0}^{+\infty} \frac{x^n}{E(n)!}.
\label{eexp}
\ee 
Then, using (\ref{state1}), the  coherent state (\ref{cs1}) and the normalization constant (\ref{cs2}) can be written in simpler form, respectively $\vert z_E \rangle =  e_E  ^{z a_E^{\dagger}} \vert 0 \rangle$ and ${\cal N}_E (z) = (e_E^{\vert z \vert^2} )^{-1/2}$. 

$\bullet$ For the identity function $E=id$ we have $e^x_{id} = e^x$ as an appropriate limit, see Eq.~(\ref{A4}) of the Appendix. Then, the vectors $\vert z_{id} \rangle_N \equiv \vert z \rangle_N$ represent the conventional coherent states of the harmonic oscillator.

%---------------------------------------> Subsection
\subsection{Completeness}

As usual, although the nonlinear coherent states $\vert z_E \rangle_N$ are not mutually orthogonal $\langle z_E \vert z' _E \rangle = {\cal N}_E(z) {\cal N}_E(z') e^{z^* z'}_E$ (hereafter the symbol~${}^*$ denotes complex conjugation), they satisfy a closure relation
\be
\mathbb I = \int {\cal N}_E^{\, 2} (z)  \vert z_E \rangle \langle z_E \vert \, d\mu_E (z),
\label{closure}
\ee
with $d\mu_E (z)$ a measure function to be determined. Let us write 
\be
d\sigma_E (z) = {\cal N}_E^{\, 2} (z) \, d\mu_E (z) = \frac{d^2 z}{\pi} \Lambda_E (\vert z \vert^2),
\label{meas1}
\ee
where $\Lambda_E$ is an additional function to be determined, $d^2 z= r dr d\theta$, and $z= r e^{i\theta}$. After integrating over $\theta$, the expression (\ref{closure}) is as follows
\be
\mathbb I = \sum_{n=0}^{+\infty} \frac{ \vert n \rangle \langle n \vert}{E(n)!} \int_0^{\infty} \Lambda_E (x) x^n dx, \quad x=r^2.
\ee
As the number eigenvectors $\vert n \rangle$ form a complete set, the above equality is achieved whenever $\Lambda_E$ satisfies 
\be
\int_0^{\infty} \Lambda_E (x) x^{n} dx = E(n)!
\label{moment}
\ee
After the change $n \rightarrow m-1$, the integral equation (\ref{moment}) coincides with the Mellin transform \cite{Ber00} of $\Lambda_E(x)$. 

%---------------------------------------> Subsection
\subsection{Examples}
\label{Ch3examp}

Once the algebras (\ref{alg2})--(\ref{alg3}) and the related coherent states (\ref{cs1}), with closure relation (\ref{closure}), have been given in general form, it is profitable to analyze concrete realizations in detail. The coherent states for the $f$-oscillators of Sec.~\ref{fosc} and those for the $q$-oscillators of Sec.~\ref{qosc} have been exahustively studied in \cite{Man97} and \cite{Bie89,Mac89}, respectively. Hence, it is not necessary to revisit them in the present work. Nevertheless, we would like to mention that the former have been used in approaching the Jaynes-Cummings model for some nonlinear Kerr media \cite{San12}, and that the coherent states of the $q$-oscillators can be recovered from those reported in \cite{Man97} as a particular case (see \cite{Ari76,Jan81} for early constructions using other definition of the $E$-function). Next, we pay attention to the coherent states associated with the polynomial algebras derived in Sec.~\ref{polosc}. The reason is that such algebras are general enough to include a plenty of cases which have a common property. Namely, the measure permitting the resolution of the identity (\ref{closure}) is given in terms of a Meijer $G$-function.  

%---------------------------------------> Subsubsection
\subsubsection{Polynomial-like oscillators}

The explicit form of the $E$-exponential function (\ref{eexp}) for the $\ell$-polynomial $E$-function (\ref{epol1})-(\ref{epol2}) is given in Eq.~(\ref{A2}) of the Appendix. Except for some atypical cases, the normalization constant (\ref{cs2}) is well defined so that the related coherent states $\vert z_E \rangle_N$ are in the Hilbert space ${\cal H}$. To satisfy the closure relation (\ref{closure}), in this case the Mellin transform (\ref{moment}) is simplified by using the change of variables:
\be
y = \frac{x}{\gamma_{\ell}}, \quad M_E(y) = \left[ \prod_{p=1}^{\ell} \Gamma(1+\delta_p) \right] \gamma_{\ell} \Lambda_E(x).
\label{meas2}
\ee
Thus, we arrive at the moment problem 
\be
\int_0^{\infty} M_E(y) y^{n-1} dy= \Gamma (n+\delta_1) \cdots \Gamma(n+\delta_{\ell}),
\label{mellin}
\ee
which is the Mellin-Barnes integral representation \cite{Olv10} of the following Meijer G-function 
\be
M_E(y) = G_{0,\ell}^{\ell,0} \left(
\begin{array}{c|c}
y & \begin{array}{c}
 -\\[1ex]
\delta_1, \cdots, \delta_{\ell}
\end{array}
\end{array}
\right).
\label{meas3}
\ee
After substituting this last result into (\ref{meas2}) we obtain the explicit form of the measure (\ref{meas1}) we are looking for.

$\bullet$ For $\ell=1$, the introduction of $E(n)= \alpha n + \beta$ in (\ref{cs1})--(\ref{cs2}) gives the normalized vectors
\be
\vert z_E \rangle_N = \left[ \frac{\Gamma(1+\delta)}{ {}_1F_1 (1,1+\delta, \vert z \vert^2/\alpha) } \right]^{1/2} \sum_{n=0}^{+\infty} \frac{ (z/\sqrt \alpha)^n}{ \sqrt{ \Gamma(n+1+\delta)}} \vert z \rangle,
\label{csele1}
\ee
where we have used the Eq.~(\ref{A3}) of the Appendix. The Meijer G-function (\ref{meas3}) is very simple in this case $G_{0,1}^{1,0} (x, \delta) = e^{-x} x^{\delta}$. Then, the nonlinear coherent states (\ref{csele1}) form an over-complete set in the Hilbert space ${\cal H}$ whenever $E(0)=0$, see paragraph between Eqs.~(\ref{alg3}) and (\ref{state1}). Then $\alpha=1$ and $\beta=0$ (equivalently $\delta =0$), so that (\ref{csele1}) is reduced to the expression of the conventional coherent states $\vert z_{id} \rangle_N$.

%---------------------------------------> Subsubsection
\subsubsection{$su(1,1)$-oscillators}
\label{Secsu11}

For the $E$-function derived in Sec.~\ref{S_su11}, the explicit form of $e^x_E$ is given in Eq.~(\ref{A5}) of the Appendix. In particular, using $\alpha_1=\alpha_2 =\beta_1=1$ one gets the expression $e^x_E=  I_1 (2\sqrt{x})/\sqrt{x}$, with $I_{\nu}(z)$ the modified Bessel function of the first kind \cite{Olv10}. Then, the normalization constant ${\cal N}_E(z) = \sqrt{ \vert z \vert / I_1(2 \vert z \vert)}$ is finite for any $z\in \mathbb C$, and the measure acquires the form
\be
d\mu_E(z)= \frac{d^2 z}{\pi} \frac{I_1(2\vert z \vert)}{\vert z \vert} G_{0,2}^{2,0} \left(
\begin{array}{c|c}
\vert z \vert^2 & \begin{array}{c}
 -\\[1ex]
1, 0
\end{array}
\end{array}
\right) 
=  \frac{d^2 z}{\pi}  2 I_1(2\vert z \vert) K_1(2 \vert z \vert),
\ee
with $K_{\nu}(z)$ the modified Bessel function of the second kind \cite{Olv10}. Therefore, the generalized $SU(1,1)$ coherent states
\be
\vert z_E \rangle_N= \left[ \frac{ {\vert z \vert} }{ I_1(2\vert z \vert) } \right]^{1/2}
\sum_{n=0}^{+\infty} \frac{z^n}{ \sqrt{ n! (n+1)! } } \vert n \rangle
\label{csimp}
\ee
form an over-complete set in the Hilbert space ${\cal H}$. 

\begin{figure}[htb]

\centering
\subfigure[ ]{\includegraphics[width=0.25\textwidth]{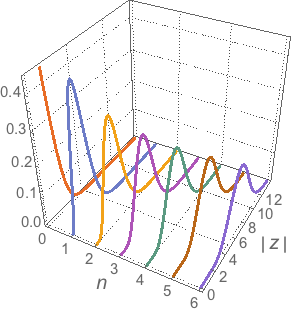} } 
\hskip4ex
\subfigure[ ]{\includegraphics[width=0.3\textwidth]{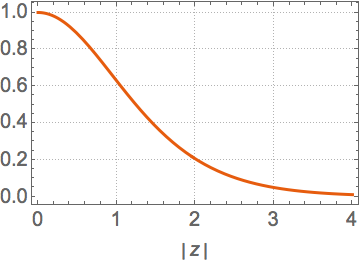} } 
\hskip4ex
\subfigure[ ]{\includegraphics[width=0.3\textwidth]{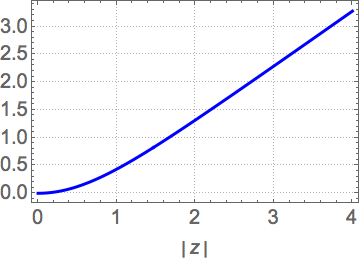} } 

\caption{\footnotesize (a) The probability ${\cal P}_E (n, \vert z \vert)$ of detecting $n$ photons associated with the $SU(1,1)$ coherent states (\ref{csimp}) (b) The probability ${\cal P}_E (0, \vert z \vert)$  of finding the system in the vacuum state as a function of $\vert z \vert$ (c) The average photon number $\langle \hat n \rangle_{z_E}$ in terms of $\vert z \vert$.
}
\label{f_prob}
\end{figure}

The probability ${\cal P}_E (n, \vert z \vert)$ of detecting $n$ photons and the average photon number $\langle \hat n \rangle_{z_E} $ are in this case given by the expressions
\be
{\cal P}_E (n, \vert z \vert)=\frac{1}{\Gamma(n+1) \Gamma(n+2) } \frac{ \vert z \vert^{2n+1} }{I_1(2 \vert z \vert)}, \qquad 
\langle \hat n \rangle_{z_E} = \frac{I_2( 2 \vert z \vert)}{ I_1 (2 \vert z \vert) } \vert z \vert.
\label{nimp}
\ee
In Fig.~\ref{f_prob}(a) we can appreciate that the maximum of the probability ${\cal P}_E (n, \vert z \vert)$ is shifted to the right of the $\vert z \vert$-axis as $n \rightarrow \infty$. For $n=0$, the probability decreases exponentially as $\vert z \vert \rightarrow \infty$, see Fig.~\ref{f_prob}(b). The latter is consistent with the behavior of the average photon number $\langle \hat n \rangle_{z_E}$ since it grows up linearly with $\vert z \vert$, see Fig.~\ref{f_prob}(c).

%%%%%%%%%%%%%
%%%%%%%%%%%

%---------------------------------------> Subsubsection
\subsubsection{SUSY-oscillators}

The nonlinear coherent states for the supersymmetric oscillators of Secs.~\ref{secsusy} and \ref{secdist} are constructed by adjusting the superpositions (\ref{cs1}) to the rule (\ref{susy3}) since the vacuum $\vert 0 \rangle$ and the $1$-photon $\vert 1 \rangle$ states are annihilated by both ladder generators, $\hat a_E$ and $\hat a_E^{\dagger}$, just as this has been indicated in Eq.~(\ref{susy1}). They can be also obtained as a limit case from either the Hermitian approaches reported in \cite{Fer95,Ros96} (important improvements are reported in \cite{ Fer07}), or the non-Hermitian ones introduced in \cite{Ros17,Zel17}. 

%---------------------------------------> Section
\section{Hilbert spaces of analytic functions}
\label{hilbert}

One of the main properties of the coherent states is their ability to form a basis of the Hilbert space ${\cal H}$, even when they are nonorthogonal \cite{Gla07,Kla68}. In this form, they can be used to represent not only any vector in ${\cal H}$, but also the operators defined to act on ${\cal H}$ in closed form \cite{Gla63,Sud63}. In the previous section we have shown that the coherent states (\ref{cs1}) associated with the generalized oscillator algebras (\ref{alg2})-(\ref{alg3}) satisfy the identity resolution (\ref{closure})-(\ref{meas1}). Here we shall go a step further by applying such property in constructing new representation spaces for the states of the new oscillators.

Using the identity resolution (\ref{closure})-(\ref{meas1}), the vectors (\ref{state2}) can be expressed as a superposition of coherent states
\be
\vert \psi \rangle = \int d\sigma_E (z) \psi(z) \vert z^*_E \rangle, 
\ee
where $d\sigma_E(z^*)= d\sigma_E (z)$, and the complex series
\be
\psi(z) :=\langle z^*_E \vert \psi \rangle = \sum_{n=0}^{+\infty} \frac{z^n \psi_n}{\sqrt{ E(n)! }}
\label{rep1}
\ee
defines the representation of $\vert \psi \rangle$ in the basis $\vert z_E \rangle_N$. As we are assuming that $E(n)$ is such that ${\cal N}_E(\vert z \vert)$ in (\ref{cs2}) is finite, it could be shown that (\ref{rep1}) converges for all finite $\vert z \vert$. Concrete realizations depend on the explicit form of the $E$-function. In particular, for the cases discussed in Sec.~\ref{Ch3examp}, the $\psi(z)$ are complex-valued functions which are analytic over the whole complex $z$-plane \cite{Ros96,Ros17}. Indeed, as these functions are holomorphic and are in one-to-one correspondence with the number eigenstates, they are elements of a Hilbert space of entire functions ${\cal F}_E$ named after Fock \cite{Foc28} and Bargmann \cite{Bar61}. In general, from the Schwarz inequality we get $\vert \psi(z) \vert \leq {\cal N}_E^{-1}(z) \vert \vert \, \vert \psi \rangle \vert \vert$, so that the growth of $\vert \psi(z) \vert$ will be bounded from above by the reciprocal of the normalization constant. In such a representation it follows that
\be
a_{E, op}^{\dagger} \psi(z) := \langle z^*_E \vert a^{\dagger}_E \vert \psi \rangle = \sum_{n=0}^{+\infty} \frac{z^{n+1} \psi_n}{\sqrt{ E(n)! }} = z \psi(z).
\ee
Thus, the action of the creation operator $\hat a_E^{\dagger}$ on the space ${\cal F}_E$ is reduced to the multiplication by $z$. On the other hand, for the annihilation operator one gets
\be
a_{E, op} \psi(z) : = \langle z^*_E \vert a_E \vert \psi \rangle = \sum_{n=1}^{+\infty} \frac{z^{n-1}}{\sqrt{ E(n)! }} E(n) \psi_n.
\ee
Comparing with the derivative of $\psi(z)$ with respect to $z$,
\be
\frac{d}{dz} \psi(z) = \sum_{n=1}^{+\infty} \frac{n z^{n-1}}{\sqrt{ E(n)!}} \psi_n,
\label{bop}
\ee
we realize that $\hat a_E$ is not the canonical conjugate of $\hat a_E^{\dagger}$ in ${\cal F}_E$ for arbitrary forms of $E$. Nevertheless, the operator $\hat b_E$ that corresponds to (\ref{bop}) and satisfies $[\hat b_E, \hat a_E^{\dagger}]=1$ in ${\cal F}_E$ produces also the linearization of the algebra (\ref{alg1})-(\ref{alg3}). Preliminary results on the matter can be found in \cite{Fer07}, the detailed construction of $\hat b_E$ for the general case we are dealing with will be reported elsewhere.

Of course, if $E=id$ the above expressions are reduced to those of the conventional oscillator for which the Fock-Bargmann space ${\cal F}_{id}$ is formed by entire analytic functions of growth $(\sfrac12,2)$. In this case, the usual boson operators $\hat a_{id}^{\dagger} = \hat a^{\dagger}$ and $\hat a_{id} = \hat a$ correspond to the multiplication by $z$ and the derivative with respect to $z$, respectively. 

The representation (\ref{rep1}) is useful to describe pure states only. A more general and versatile representation is offered by the density operator $\hat \rho$ which includes pure states, $\mbox{Tr} \hat \rho^2 = \mbox{Tr} \hat \rho$, as well as mixed states $\mbox{Tr} \hat \rho^2 < \mbox{Tr} \hat \rho$. Following \cite{Gla63,Sud63}, let us write $\hat \rho$ in $P$-representation
\be
\hat \rho = \int d\sigma_E (z) P(z) \vert z_E \rangle \langle z_E \vert.
\label{P1}
\ee
The main point here is to find the appropriate $P$-function such that (\ref{P1}) can be interpreted as a `diagonal' continuous matrix representation of $\hat \rho$. In doing so, $P(z)$ would play the role of a nonnegative weight function, defined at all points of the complex $z$-plane. 

As indicated earlier, for $E={id}$ we recover the conventional coherent states  $\vert z_{id} \rangle_N =\vert z \rangle_N$ of the harmonic oscillator. In such a case the density operators (\ref{P1}) describe the light emitted by a completely chaotic source, a model that includes all known natural light sources \cite{Gla07}. Also in this case the $P$-function need not have the properties of a probability distribution \cite{Sud63} and does not exist for all $\hat \rho$ \cite{Gla63}. However, as the classical probability theory allows for delta function distributions, $P$ can be as singular as $\delta^{(2)} (z) = \delta ( \mbox{Re} z)  \delta ( \mbox{Im} z)$ \cite{Sud63,Gla63}. In the quantum theory of optical coherence \cite{Gla07}, if the $P$-function of a given state $\hat \rho$ doest not possess properties of a classical probability distribution, or it does not exist, such state does not have classical analog. Coming back to our approach, in the simplest case, the density operator $\hat \rho_z = \vert z_E \rangle \langle z_E \vert$ of any of the coherent states in the sum (\ref{P1}) should be represented by a distribution $\delta^{(2)}(z -z')$. Otherwise, the  identity resolution (\ref{closure})-(\ref{meas1}) would be not valid. Thus, as in the conventional case, the function $P(z)$ should have, at most, $\delta$-type singularities.

To investigate the $P$-representation for other states consider the superposition
\be
\vert \beta \rangle = \sum_{n=0}^{+\infty} \frac{\beta^n \sqrt{E(n)!}}{n!} \vert n \rangle,
\ee
we may calculate the matrix elements
\be
\langle -\beta \vert \hat \rho \vert \beta \rangle = \int d\sigma_E(z) P(z) e^{\beta z^* -\beta^* z}.
\ee
Then, the $P$-function is obtained from the two-dimensional inverse Fourier transform 
\be
P(z)= \frac{1}{\Lambda(r^2)} \int d^2 \beta  \langle -\beta \vert \hat \rho \vert \beta \rangle e^{\beta^* z -\beta z^*}.
\ee
As an immediate application consider the number eigenstate $\hat \rho_n = \vert n \rangle \langle n \vert$, then
\be
P_n(z)= \frac{E(n)!}{(n!)^2\Lambda(r^2)} \frac{ \partial^{2n}}{ \partial z^n \partial z^{*  n}} \delta^{(2)}(z), \quad n\geq 0.
\ee
That is, with exception of the vacuum $\vert 0 \rangle$, the $P$-representation of the number eigenstates $\vert n \rangle$ is as singular as the derivatives of the $\delta$-distribution. Therefore, according to \cite{Gla07}, the fields represented by any of the number eigenstates $\vert n+1 \rangle$ do not have classical analog. This result is quite natural since the states $\vert n+1 \rangle$ cannot be described in classical terms, no matter the approach used in their study. In this form, the continuous matrix representation (\ref{P1}) is consistent with the results obtained in terms of the conventional coherent states for the vectors $\vert n+1 \rangle$.

On the other hand, the $P$-representation of the vacuum $\vert 0 \rangle$ and the generalized coherent states $\vert z_E \rangle_N$ is the delta function $\delta^{(2)}(z-z')$, with $z' =0$ for $\vert 0 \rangle$. The above criterion of classicality would mean that the nonlinear coherent states $\vert z_E \rangle_N$ are able to represent fields with classical analog, at least at the same level as the vacuum $\vert 0 \rangle$. The latter because the singularity of the delta distribution is integrable, so that $P=\delta^{(2)}$ is admissible as a classical probability distribution \cite{Gla07}. Such statement is rather clear for $E=id$ since the conventional coherent states $\vert z_{id} \rangle_N$ are indeed as classical as the vacuum $\vert 0 \rangle$. The situation changes if $E\neq id$, as we are going to see in the next sections.

%---------------------------------------> Section
\section{Do the completeness of generalized coherent states imply classicality?}
\label{fundamento}

We have shown that all the nonlinear coherent states $\vert z_E \rangle_N$ associated with the $\ell$--oder polynomials (\ref{epol1})-(\ref{epol2}) have a resolution to the identity (\ref{closure}) that is expressed in terms of the Meijer $G$-function (\ref{meas3}). The latter means that the $P$-representation of all the superpositions  (\ref{cs1}) is the distribution $\delta^{(2)}(z-z')$, just as this occurs for the vacuum $\vert 0 \rangle$ which is $P$-represented by $\delta^{(2)}(z)$. Hence, $\vert z_E \rangle_N$ is a displaced version of $\vert 0 \rangle$. The question is if such property is a sufficient condition for the vectors $\vert z_E \rangle_N$ to represent fields with classical analog. We look for an answer to this problem by using the criterion introduced in \cite{Kim02}, as well as the parameter introduced by Mandel \cite{Man79}, to identify the possible classicality of such states.

%---------------------------------------> Subsection
\subsection{Beam splitter criterion}
\label{divisor}

\begin{figure}[htb]

\centering
\includegraphics[width=0.3\textwidth]{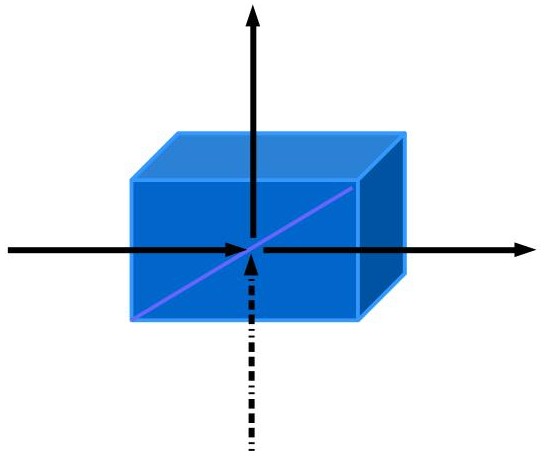} 

\caption{\footnotesize The $50\!:\!50$ beam splitter represented by the operator $BS$ in (\ref{bsop}). This is a two-channel optical device operating on states $\vert \cdot \rangle \otimes \vert \cdot \rangle$, where the ket at the left (right) stands for `horizontal' (`vertical') signal. The dotted arrow represents classical states like the vacuum $\vert 0 \rangle$ while the continuous arrows are associated with nonclassical states like the number eigenstates $\vert n+1 \rangle$. 
}
\label{f_BS}
\end{figure}

Consider a single photon $\vert 1 \rangle$ entering a $50\!:\!50$ beam splitter
\be
BS= \exp\left[ i \frac{\pi}{4} \left(a^{\dagger}_H a_V + a_H a^{\dagger}_V \right) \right].
\label{bsop}
\ee
The subindex `$H$'  (`$V$') stands for the horizontal (vertical) channel of the beam splitter, see Fig.~\ref{f_BS}. As a vacuum $\vert 0 \rangle$ enters the other input port of the beam splitter, the entire input state is the product $\vert  1 \rangle \otimes \vert 0 \rangle \equiv \vert 1,0 \rangle$. Hereafter the ket at the left (right) in the tensor products $\vert \cdot \rangle \otimes \vert \cdot \rangle$ stands for `horizontal' (`vertical') signal with respect to the beam splitter shown in Fig.~\ref{f_BS}. The output is the Bell state $ \vert \beta \rangle = \frac{1}{\sqrt 2} (\vert 1,0 \rangle +i  \vert 0,1 \rangle)$ which, as it is well known, encodes nonclassical correlations \cite{Bou00,Sil08}. Such state is distinguished from the classical correlation $\rho_{clas} = \frac12 (\vert 1 \rangle \langle 1 \vert \otimes \vert 0 \rangle \langle 0 \vert + \vert 0 \rangle \langle 0 \vert \otimes \vert 1 \rangle \langle 1 \vert)$ because the off-diagonal elements of its density operator $\vert \beta \rangle \langle \beta \vert$ are associated with transitions $\vert  0 \rangle \leftrightarrow \vert 1 \rangle$, occurring in both channels, that are invariant under a change of basis. That is, the off-diagonal elements $\vert \beta \rangle \langle \beta \vert - \rho_{clas}$ that are different from zero produce entanglement. In general, when the state $\vert n,0 \rangle$ enters the beam splitter, one gets the well known binomial distribution of bi-partite photon states
\be
BS \vert n, 0 \rangle=  \frac{1}{2^{n/2}} \sum_{k=0}^n  \left(
\begin{array}{c} n\\ k
\end{array}
\right)^{1/2}  e^{i\frac{\pi}{2} k} \vert k, n-k \rangle,
\label{bin}
\ee
where a global phase has been dropped. The straightforward calculation shows that the off-diagonal elements of the density operator $BS \vert n, 0 \rangle \langle n,0 \vert (BS)^{\dagger}$ are different from zero, so that the pure state (\ref{bin}) encodes nonclassical correlations. These off-diagonal elements are such that measuring the number of photons at the horizontal output port of the beam splitter is affected by the result of detecting photons at the vertical port and vice versa. This last result has motivated the conjecture that the entangled output state from a beam splitter requires nonclassicality in the input state \cite{Kim02}. Assuming that the conjecture can be proved (see, e.g. \cite{Wan02}), this would be used as a criterion for nonclassicality. For if the off-diagonal terms of the output state $\vert \psi_{out} \rangle \langle \psi_{out} \vert$ are nontrivial, then the input state $\vert \psi_{in} \rangle \langle \psi_{in} \vert$ is nonclassical in at least one of its two channels \cite{Kim02}. 

%%%%%%%%%%%%%
%%%%%%%%%%%

Let us analyze the presence of nonclassical correlations in (\ref{bin}) by considering the probability of finding $m$ and $r$ photons in the horizontal and vertical channels respectively,
\be
\vert \langle m, r \vert BS \vert n, 0 \rangle \vert^2=   \frac{ \Gamma(r+\frac12) \Gamma(m+\frac12)}{\Gamma(r+1) \Gamma(m+1)} \left[ \frac{1}{ 2^{r+m} B(r+ \frac12, m+ \frac12 )}  \right].
\label{beamp}
\ee
Here, $B(a,b)$ stands for the Euler beta function \cite{Olv10} which,  as far as we know, cannot be expressed as $B(a,b) = f(a) g(b)$ for any functions $f$ and $g$. Therefore, the probability (\ref{beamp}) cannot be factorized as the product of two independent distributions, one for each output port of the beam splitter. This property is concomitant to the impossibility of writing the bi-partite photon states (\ref{bin}) as the product of any state $\vert \phi_H \rangle$ of the horizontal channel with a state $\vert \phi_V \rangle$ of the vertical one, that is $BS \vert n, 0 \rangle \neq \vert \phi_H \rangle \otimes \vert \phi_V \rangle$ if $n \neq 0$.

On the other hand, a measure of the nonclassicality of states is given by the Mandel parameter 
\be
Q = \frac{ (\Delta n)^2}{ \langle \hat n \rangle} -1,
\label{Mand}
\ee
which indeed indicates the degree to which the statistics of a given field is sub-Poissonian \cite{Man79}. For a field represented by the number eigenvector $\vert n +1 \rangle$ one gets $Q=-1$, so that the field is sub-Poissonian ($-1 \leq Q <0$). Classical fields like those represented by either a coherent state or a vacuum are Poissonian ($Q=0$), or even super-Poissonian ($Q>0$) if they correspond to  thermal light. 

Now, let us write $\hat n_{tot} = \hat n_H +\hat n_V$ for the total number operator associated to either the input or the output ports of the beam splitter. Here $\hat n_H = \hat n \otimes \mathbb I$ and $\hat n_V= \mathbb I \otimes \hat n$ are the number operators for the horizontal and vertical channels respectively. We may prove the expression $(\Delta n_{tot})^2 =(\Delta n_H)^2 + (\Delta n_V)^2$, meaning that the variance of any signal involved with a beam splitter is the result of adding the variances of the horizontal and vertical channels. A simple calculation shows that (\ref{bin}) is such that $\langle \hat n_H \rangle = \langle \hat n_V \rangle =\frac{n}{2}$, $\langle \hat n_H^2 \rangle = \langle \hat n_V^2 \rangle =\frac14 n(1+n)$, so that $Q_H=Q_V=-\frac12$. Thus, the horizontal and vertical signals of (\ref{bin}) are sub-Poissonian (nonclassical).

For the present case, the conjecture indicated in \cite{Kim02}, hereafter the ({\em Knight\/}) K-conjecture, is trivially  verified since the number eigenstate $\vert n \rangle$ injected into the beam splitter to produce the superposition (\ref{bin}) is clearly nonclassical if $n\neq 0$.

%---------------------------------------> Subsection
\subsection{Classical signals}

\begin{figure}[htb]

\centering
\subfigure[ ]{\includegraphics[width=0.3\textwidth]{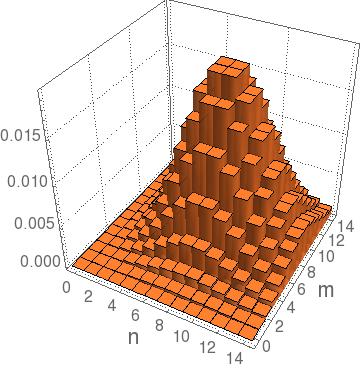} } 
\hskip2ex
\subfigure[ ]{\includegraphics[width=0.3\textwidth]{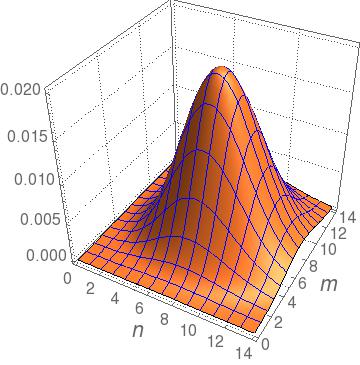} } 
\hskip2ex
\subfigure[ ]{\includegraphics[width=0.3\textwidth]{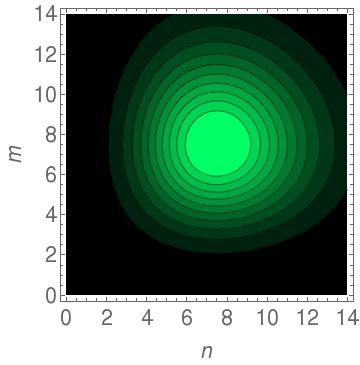} } 

\caption{\footnotesize Using the classical state $\vert z, 0\rangle$ as input in the beam splitter of Fig.~\ref{f_BS}, the probability ${\cal P}(n,m,\vert z \vert)$ of detecting $n$ and $m$ photons at the horizontal and vertical output ports is factorizable as the product of two independent Poisson distributions (\ref{poisson}). The figures (a) and (b) correspond to the distribution ${\cal P}(n,m,4)$ obtained for an input signal with $\langle \hat n \rangle =16$, and (c) shows some of its level curves respectively.
}
\label{f_G1}
\end{figure}

\begin{figure}[htb]

\centering
\subfigure[ $\vert z \vert =1 $]{\includegraphics[width=0.2\textwidth]{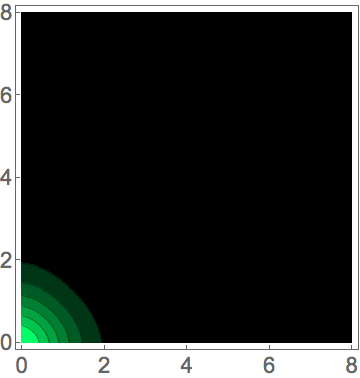} } 
\hskip4ex
\subfigure[  $\vert z \vert = 1.25$]{\includegraphics[width=0.2\textwidth]{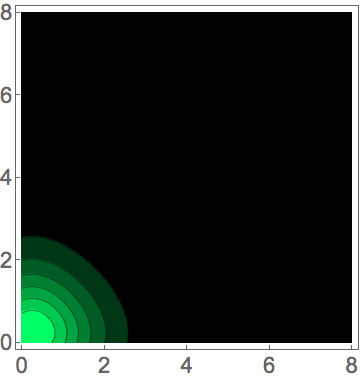} } 
\hskip4ex
\subfigure[  $\vert z \vert = 1.5$]{\includegraphics[width=0.2\textwidth]{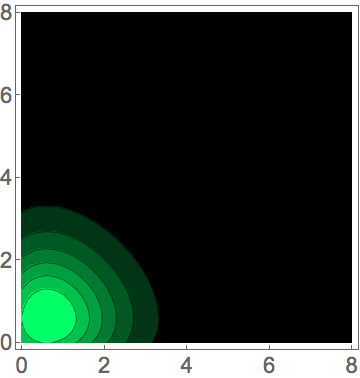} } 
\hskip4ex
\subfigure[ $\vert z \vert = 1.75$ ]{\includegraphics[width=0.2\textwidth]{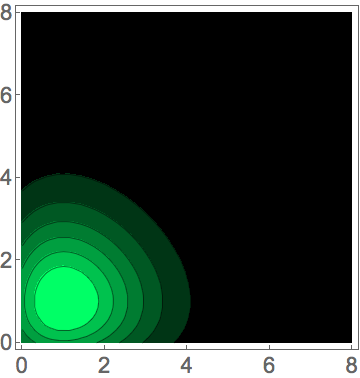} } 
\vskip1ex
\centering
\subfigure[ $\vert z \vert = 2$]{\includegraphics[width=0.2\textwidth]{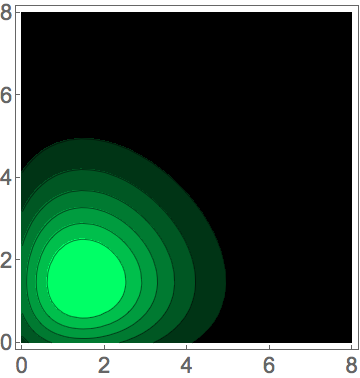} } 
\hskip4ex
\subfigure[  $\vert z \vert = 2.25$]{\includegraphics[width=0.2\textwidth]{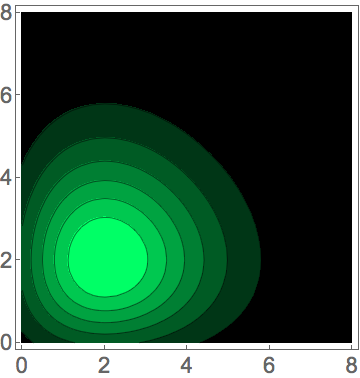} } 
\hskip4ex
\subfigure[  $\vert z \vert = 2.5$]{\includegraphics[width=0.2\textwidth]{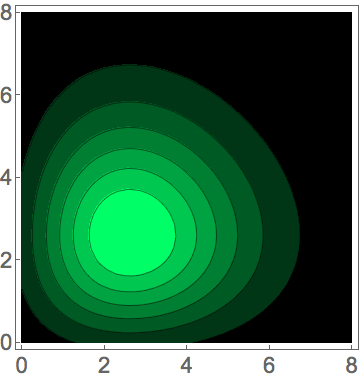} } 
\hskip4ex
\subfigure[ $\vert z \vert = 2.75$ ]{\includegraphics[width=0.2\textwidth]{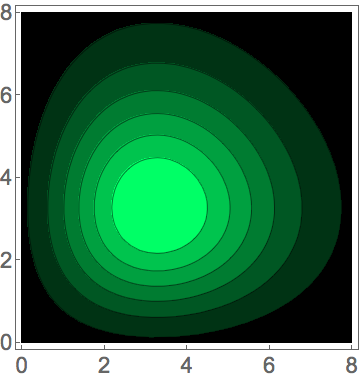} } 

\caption{\footnotesize The distribution ${\cal P}(n,m,\vert z \vert)$ in the $nm$-plane for the indicated values of $\vert z \vert$. 
}
\label{f_G2}
\end{figure}

Although the above discussion is true for the number eigenvectors $\vert n+1 \rangle$ in any of the input channels, the result cannot be generalized for any superposition of such states. For instance, the input state $\vert z, 0 \rangle$, with $\vert z_{id} \rangle_N =\vert z\rangle_N$ a conventional coherent state of the harmonic oscillator, produces the separable (classical) signal
\be
BS \vert z,0 \rangle= \vert z/\sqrt{2} \rangle_N \otimes \vert i z/\sqrt{2} \rangle_N.
\label{bsg}
\ee
The latter confirms that the coherent state $\vert z \rangle_N$ is classical. The separability of the state $BS \vert z, 0 \rangle$ can be also studied in terms of the probability ${\cal P}(n,m,\vert z \vert) =  \vert \langle n,m \vert BS \vert z, 0 \rangle \vert^2$ of detecting $n$ photons in the horizontal channel and $m$ photons in the vertical one. The result is shown in Fig.~\ref{f_G1} for $\langle \hat n \rangle =16$ in the input signal. In general, the distribution spreads out on the $nm$-plane while its center is shifted along the line $m=n$ as $\vert z \vert \rightarrow \infty$, see Fig.~\ref{f_G2}.

%%%%%%%%%%%%%
%%%%%%%%%%%

In this case the probability ${\cal P}(n,m,\vert z \vert)$ can be expressed as the product of two Poisson distributions with mean value $\vert z \vert^2/\sqrt{2}$, one for each output port,
\be
P_{oisson} \left( \frac{\vert z \vert }{ \sqrt{2}} , n \right) = \frac{e^{-\vert z \vert^2/2} }{\Gamma(n+1)} \left( \frac{\vert z \vert^2}{2} \right)^n.
\label{poisson}
\ee
The latter means that measuring the number of photons at the horizontal output port does not depend on the result of detecting photons at the vertical output port of the beam splitter. This result enforces the notion of classicality associated to the conventional coherent states of light.

%%%%%%%%%%%%%
%%%%%%%%%%%

Now, it can be shown that the state (\ref{bsg}) gives $(\Delta n_{\ell} )^2_{out}  =\langle \hat n_{\ell} \rangle_{out}$, with $\ell = H,V$. Then $Q_{H,out}=Q_{V,out}=0$. The latter is consistent with the separability of ${\cal P}(n,m,\vert z \vert)$ indicated above since the same Poisson distribution (\ref{poisson}) determines the photon-detection for the two output ports of the beam splitter.  

Summarizing the properties of the state (\ref{bsg}), generated when the classical signal $\vert z_{id} \rangle_N$ enters a 50:50 beam splitter, we have

(1.C) The average occupation number ${\cal P}(n,m,\vert z \vert)$ can be factorized as the product of two independent Poisson distributions, one for each output port.

(2.C) The variances of the input and output signals are equal.

(3.C) The variances of the horizontal and vertical output signals coincide and are equal to one half the variance of the input signal.

The point (1.C) and the product in Eq.~(\ref{bsg}) are concomitant since we can take one of them as a given property to verify the other one, and vice versa. Here, we would like to emphasize that the points (2.C) and (3.C) would serve as a criterion to investigate the possibility of factorizing the state $BS \vert \psi, 0 \rangle$, equivalently ${\cal P}(n,m,\vert z \vert)$, when $\vert \psi \rangle$ is an arbitrary superposition of photon states. As the factorization of  $BS \vert \psi, 0 \rangle$ means no quantum correlations, the properties (2.C) and (3.C) may imply the classicality of $\vert \psi \rangle$.

As we can see, also in this case the K-conjecture is trivially  verified since the conventional coherent state $\vert z_{id} \rangle_N$ injected into the beam splitter to produce the vector (\ref{bsg}) is classical for any $\vert z \vert$.

%---------------------------------------> Subsection
\subsection{Nonclassical signals associated with generalized oscillators}

The above analysis can be extended to any state of the form (\ref{cs1}) that enters the beam splitter in the horizontal port, together with a vacuum $\vert 0 \rangle$ in the vertical channel. As we have seen, the separability of the distribution ${\cal P}_E(n,m,\vert z \vert)$ plays an important role in the identification of classicality. In the sequel we are going to pay special attention to the $SU(1,1)$ coherent states derived in Sec.~\ref{Secsu11}. The other coherent states mentioned in Sec.~\ref{S_nonlinear} lead to similar conclusions.

%%%%%%%%%%%%%
\begin{figure}[htb]

\centering
\subfigure[ ]{\includegraphics[width=0.3\textwidth]{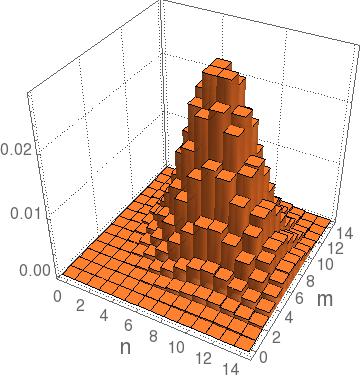} } 
\hskip1ex
\centering
\subfigure[]{\includegraphics[width=0.3\textwidth]{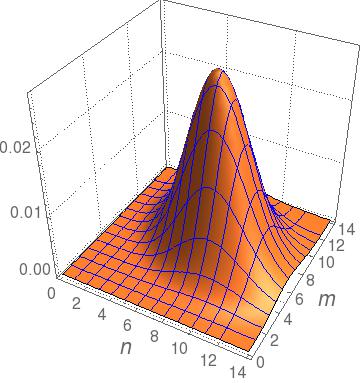} } 
\hskip1ex
\subfigure[]{\includegraphics[width=0.3\textwidth]{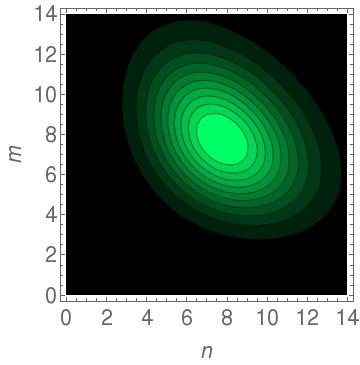} } 

\caption{\footnotesize Using the input signal $\vert z_E, 0 \rangle$, with $\vert z_E \rangle_N$ the nonlinear coherent state (\ref{csimp}) and $\langle \hat n \rangle_{z_E} =16$, the probability ${\cal P}_E(n,m,16.7442)$ of detecting $n$ and $m$ photons at the horizontal and vertical output ports of a beam splitter is shown in (a) and (b). Some of the corresponding level curves are depicted in (c). The average photon number $\langle \hat n \rangle_{z_E}$ is defined in Eq.~(\ref{nimp}) as a function of $\vert z \vert$. Compare with Fig.~\ref{f_G1}. 
}
\label{f_figL}
\end{figure}
%%%%%%%%%%%

\begin{figure}[htb]

\centering
\subfigure[ $\vert z \vert =2 $]{\includegraphics[width=0.2\textwidth]{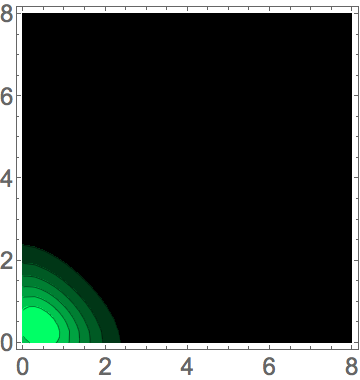} } 
\hskip4ex
\subfigure[  $\vert z \vert = 3$]{\includegraphics[width=0.2\textwidth]{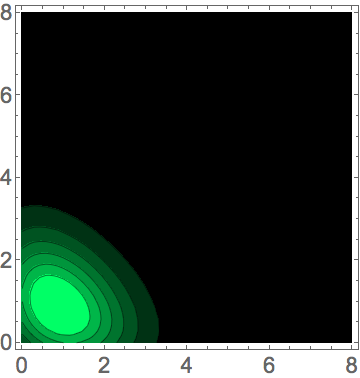} } 
\hskip4ex
\subfigure[  $\vert z \vert = 4$]{\includegraphics[width=0.2\textwidth]{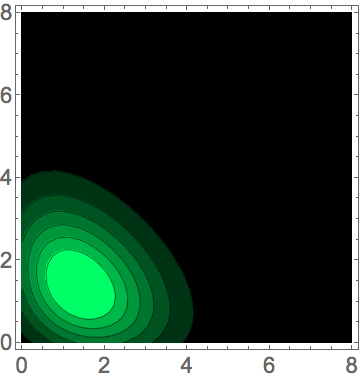} } 
\hskip4ex
\subfigure[ $\vert z \vert = 5$ ]{\includegraphics[width=0.2\textwidth]{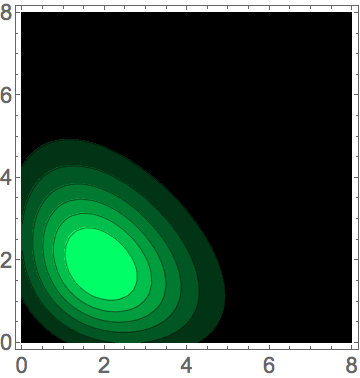} } 
\vskip1ex
\centering
\subfigure[ $\vert z \vert = 6$]{\includegraphics[width=0.2\textwidth]{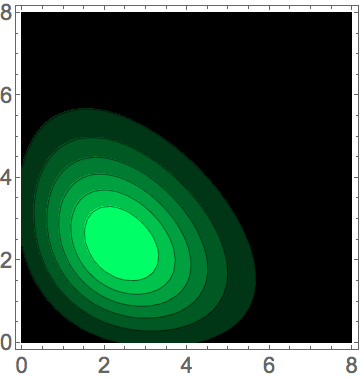} } 
\hskip4ex
\subfigure[  $\vert z \vert = 7$]{\includegraphics[width=0.2\textwidth]{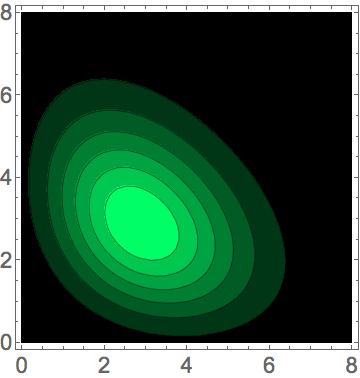} } 
\hskip4ex
\subfigure[  $\vert z \vert = 8$]{\includegraphics[width=0.2\textwidth]{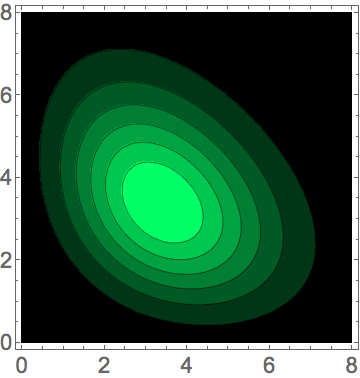} } 
\hskip4ex
\subfigure[ $\vert z \vert = 9$ ]{\includegraphics[width=0.2\textwidth]{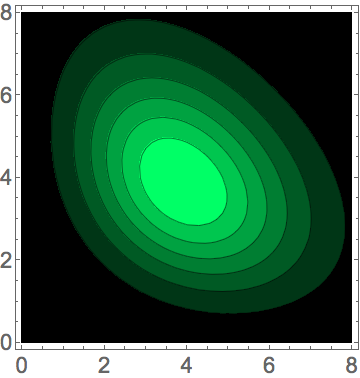} } 

\caption{\footnotesize The distribution ${\cal P}_E (n,m,\vert z \vert)$ in the $nm$-plane for the indicated values of $\vert z \vert$. Compare with Fig.~\ref{f_G2}.
}
\label{f_figm}
\end{figure}

Using the $SU(1,1)$ coherent state (\ref{csimp}), the probability of detecting $n$ photons in the horizontal channel and $m$ photons in the vertical one gives
\be
{\cal P}_E(n,m,\vert z \vert) = \frac{ 2^{-n-m} (n+1)(m+1) B(m+1,n+1) }{
\Gamma(n+1) \Gamma(n+2) \Gamma(m+1) \Gamma(m+2) I_1 (2 \vert z \vert) }
\, \vert z \vert^{2(n+m)+1}.
\label{distsu}
\ee
The behavior of this last distribution is shown in Fig.~\ref{f_figL} for the average photon number $\langle \hat n \rangle_{z_E} =16$ in the input signal, the latter value has been chosen for comparison with the result of Fig.~\ref{f_G1}. The squeezing of ${\cal P}_E(n,m,\vert z \vert)$ along the line $n=m$ is notable. Concerning the global profile, as in the previous case, the distribution (\ref{distsu}) spreads out in the $nm$-plane while its center is shifted along the line $m=n$ as $\vert z \vert \rightarrow \infty$, see Fig.~\ref{f_figm}.

%%%%%%%%%%%%%
%%%%%%%%%%%

As it occurred for the probability (\ref{beamp}), the presence of the beta function in (\ref{distsu}) prohibits the factorization of ${\cal P}_E(n,m,\vert z \vert)$ as the product of two independent probability distributions if $n$  and $m$ are both different from zero. The latter means that detecting photons at the vertical output port of the beam splitter affects the counting of photons in the horizontal output port. 

%%%%%%%%%%%%%
\begin{figure}[htb]

\centering
\subfigure[ ]{\includegraphics[width=0.3\textwidth]{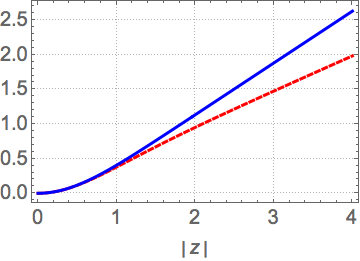} } 
\hskip4ex
\subfigure[]{\includegraphics[width=0.3\textwidth]{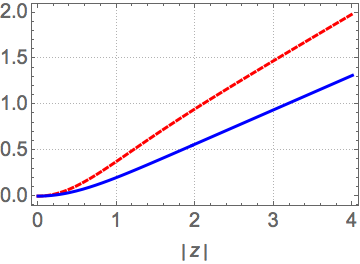} } 

\caption{\footnotesize (a) The variances of the input $\vert z_E, 0 \rangle$ and output $BS \vert z_E, 0 \rangle$ signals of a 50:50 beam splitter, red-dashed and blue respectively,  with $\vert z_E \rangle_N$ the $su(1,1)$ coherent state (\ref{csimp}) (b) The variances of the signals at the horizontal and vertical output ports are equal (blue) and shorter than the variance of the input $\vert z_E, 0 \rangle$ (red-dashed) for any $\vert z \vert \neq 0$.
 }
\label{f_variance}
\end{figure}
%%%%%%%%%%%

\begin{figure}[htb]

\centering
\subfigure[ $\vert z \vert = 4$]{\includegraphics[width=0.3\textwidth]{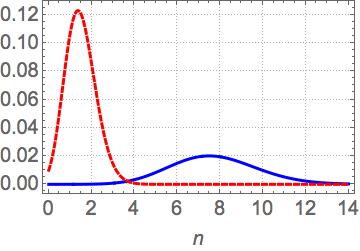} } 
\hskip4ex
\subfigure[  $\vert z \vert = 16.7442$]{\includegraphics[width=0.3\textwidth]{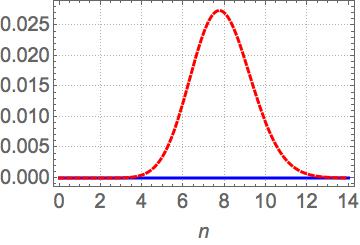} }

\caption{\footnotesize Squeezing along the line $n=m$ of the photon probability distributions ${\cal P}(n,m,\vert z \vert)$ and  ${\cal P}_E(n,m,\vert z \vert)$, blue and red-dashed respectively. In (a) the photon number averages for the incoming states $\vert z,0 \rangle$ and $\vert z_E,0 \rangle$
are respectively $\langle \hat n \rangle=16$ and $\langle \hat n \rangle_{z_E}=3.277$, compare with Fig.~\ref{f_G1}. In (b) we have $\langle \hat n \rangle=280.369$ and $\langle \hat n \rangle_{z_E}=16$, compare with Fig.~\ref{f_figL}. In all cases the maximum of ${\cal P}(n,m,\vert z \vert)$ and  ${\cal P}_E(n,m,\vert z \vert)$ is reached at $\langle \hat n \rangle/2$ and $\langle \hat n \rangle_{z_E}/2$ respectively. 
}
\label{f_comp}
\end{figure}

In turn, the variances $(\Delta n_{E, tot})^2$ and $(\Delta n_{E, tot})^2_{out}$ of the input $\vert z_E, 0 \rangle$ and output $BS \vert z_E, 0 \rangle$ signals are not the same. Indeed, the latter is larger than the former for practically any $\vert z \vert \neq 0$, see Fig.~\ref{f_variance}(a). However, the horizontal and vertical output variances coincide and they are shorter than the variance of $\vert z_E,0\rangle$, see Fig.~\ref{f_variance}(b). Such squeezing is specially clear along the line $n=m$, as this has been indicated above. See Fig.~\ref{f_comp} for a detailed comparison.

%%%%%%%%%%%%%
%%%%%%%%%%%

\begin{figure}[htb]

\centering
\includegraphics[width=0.3\textwidth]{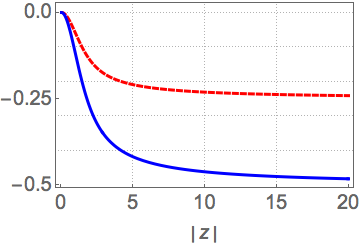} 

\caption{\footnotesize The blue curve represents the Mandel parameter $Q$ of the $SU(1,1)$ coherent states (\ref{csimp}) as a function of $\vert z \vert$. This converges to $-\sfrac12$ as $\vert z \vert \rightarrow \infty$. The dashed-red curve represents the Mandel parameter for both the horizontal and vertical signals at the output ports of a beam splitter when it is injected with the state $\vert z_E, 0 \rangle$, where $\vert z_E \rangle_N$ is the coherent state (\ref{csimp}).
}
\label{f_mandel}
\end{figure}

The above results indicate that the $SU(1,1)$ coherent sates (\ref{csimp}) are nonclassical although they can be interpreted as displaced versions of the vacuum $\vert 0 \rangle$. Indeed, the direct calculation shows that the Mandel parameter associated to $\vert z_E \rangle_N$ is such that $-\frac12 \leq Q \leq 0$ for all $\vert z \vert \geq 0$, and $Q =0$ for $\vert z \vert =0$ only. Moreover, $Q \rightarrow -\frac12$ as $\vert z \vert \rightarrow \infty$, see the blue curve depicted in Fig.~\ref{f_mandel}. That is, the statistics associated with the state (\ref{csimp}) is sub-Poissonian, so that it is nonclassical. On the other hand, the Mandel parameter for the horizontal and vertical channels of the output signal in the interferometer gives a result that is equal to one-half the result of the input signal, see the dashed-red curve in Fig.~\ref{f_mandel}. 
 
%%%%%%%%%%%%%
%%%%%%%%%%%

The properties of the state that is generated when the nonclassical signal $\vert z_E \rangle_N$ enters a 50:50 beam splitter are as follows

(1.N) The average occupation number ${\cal P}_E (n,m,\vert z \vert)$ cannot be factorized as the product of two independent probability distributions.

(2.N) The variances of the input and output signals are different.

(3.N) The variances of the horizontal and vertical output signals are shorter than one half the variance of the input signal.

The properties (1.N)--(3.N) are in opposition to the (1.C)--(3.C) ones. As the last are not satisfied, the state $BS \vert z_E \rangle$, equivalently ${\cal P}_E (n,m,\vert z \vert)$, cannot be factorized and $\vert z_E \rangle$ is nonclassical. Besides, the squeezing property (3.N) is markedly different to (3.C) and implies the squeezing of the distribution ${\cal P}_E (n,m,\vert z \vert)$ along the line $n=m$. 

Then, the K-conjecture is also verified for the nonlinear coherent states defined in (\ref{csimp}). Namely, as the output state $BS\vert z_E, 0 \rangle$ is non-separable, the state $\vert z_E \rangle_N$ in the horizontal input port of the beam splitter is nonclassical. The same conclusion is obtained for any of the nonlinear coherent states (\ref{cs1}) that can be constructed with the generalized oscillator algebras (\ref{alg1})--(\ref{alg3}). However, we have shown that these states have a $P$-representation that is proportional to $\delta^{(2)} (z-z')$, so that they can be classified as displaced versions of the vacuum $\vert 0 \rangle$. Is there any contradiction between the non-separability of the states $BS \vert z_E,0 \rangle$ and the $P$-representation of $\vert z_E \rangle_N$?

%---------------------------------------> Subsection
\subsection{Refinement of the criterion}

To clarify the results of the previous sections let us emphasize that the states $BS\vert n,0 \rangle$ defined in (\ref{bin}) are nothing but a class of generalized coherent states \cite{Per86} (see also \cite{Bar71}) associated with the $su(2)$ Lie algebra realized in terms of two oscillators, i.e., in the so-called Schwinger representation \cite{Sch65},
\be
\vert \xi \rangle = \frac{1}{(1 + \vert \xi \vert^2)^{-n/2}} \sum_{k=0}^n \left(
\begin{array}{c}
n\\
k
\end{array}
\right)^{1/2} \xi^k \vert k, n-k \rangle.
\label{sol}
\ee
A simple inspection shows that making $\xi =e^{i \frac{\pi}{2}}$ in this last expression gives the state (\ref{bin}). The states $\vert \xi \rangle$ also have a resolution to the identity and are represented by a $P$-function that is proportional to the $\delta$-distribution. However, they are nonclassical, as this has been discussed in Sec.~\ref{divisor}. 

In many ways, the action of the beam splitter on the incoming state $\vert n, 0 \rangle$ is equivalent to a double-slit interference experiment in the single-photon regime, see e.g. \cite{Gra86,Rue13}. Indeed, since the detector and electronic instrumentation dead-time limitations make difficult the direct measurement of anti-bunching in a double-slit experiment \cite{Rue13}, the output intensities are instead measured by using a 50:50 beam splitter and two detectors \cite{Gra86}. These and other photon correlation experiments have their origin in the Hanbury-Brown and Twiss experiments \cite{Han56}, the results of which have shown the importance of distinguishing between the first two orders of coherence. Following Glauber \cite{Gla07}, given the normalized form of the correlation functions
\be
g^{(n)}(x_1, \ldots,x_{2n})= \frac{G^{(n)}(x_1, \ldots,x_n)}{ \prod_{j=1}^{2n} \left\{ G^{(1)} (x_j,x_j) \right\}^{1/2} }, \quad x_j \equiv (\vec r_j, t_j),
\label{g}
\ee
the necessary condition of coherence is that $\vert g^{(j)} \vert =1$. The first-order coherence is obtained if $j=1$. This coincides with the definition of coherence used in optics previous to the theory of Glauber. On the other hand, a field characterized by $n$-th order coherence corresponds to $j \leq n$. The full coherence implies that $\vert g^{(n)} \vert =1$ holds for all $n$. Light beams from ordinary sources can be made optimally first-order coherent but they lack second-order-coherence \cite{Gla07}. In turn, the conventional coherent states are coherent to all orders. Next, we follow \cite{Gla07} and adopt the factorization of $G^{(n)}$ producing $g^{(n)}=1$ in (\ref{g}) as the definition of the $n$-th order coherent fields.

The straightforward calculation shows that the horizontal and vertical channels of the nonlinear coherent states (\ref{bin}), equivalently (\ref{sol}), lack second-order-coherence since $g^{(2)}=1- \frac{1}{n} <1$ for $n\neq 0$. This means that the photons in such ports are anticorrelated, so that they cannot be created or annihilated simultaneously. The result $g^{(2)} \neq 1$ is due to the fact that $G^{(2)}$ cannot be factorized as the product of two $G^{(1)}$--functions. The statement is equivalent to the impossibility of factorizing either the vector $BS \vert n, 0 \rangle$ or the probability (\ref{beamp}). Notice however that $g^{(2)} \rightarrow 1$ for  $n \rightarrow \infty$. That is, for a large number of photons, the state (\ref{bin}) can be associated with the results of an interference experiment involving a classical field. 

%%%%%%%%%%%%%
\begin{figure}[htb]

\centering
\includegraphics[width=0.3\textwidth]{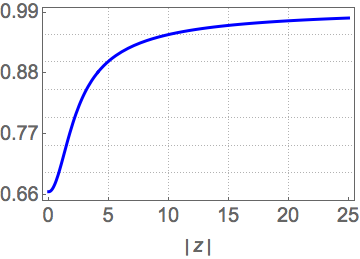} 

\caption{\footnotesize The normalized form of the correlation function of second order (\ref{g}) for the horizontal and vertical output channels of a 50:50 beam splitter injected by one of the $SU(1,1)$ coherent states $\vert z_E \rangle$ defined in (\ref{csimp}).
}
\label{f_sol}
\end{figure}
%%%%%%%%%%%

As regards the $SU(1,1)$ coherent states (\ref{csimp}), for the horizontal and vertical output ports of the beam splitter we obtain the expression 
\be
g^{(2)} =  \frac{I_1(2 \vert z \vert) I_3(2 \vert z \vert)}{I_2^2(2 \vert z\vert)} .
\ee
This last function satisfies $g^{(2)} <1$ for all finite $\vert z \vert$, and goes to one for $\vert z \vert \rightarrow \infty$, see Fig.~\ref{f_sol}.  As $\vert z \vert >>1$ implies a large average photon number, see Fig.~\ref{f_prob}(c), the latter result means that the coherent states (\ref{csimp}) allow for bunching of photons only at the limit $\langle \hat n \rangle_{z_E} \rightarrow \infty$. Again, the result $g^{(2)} \neq 1$ is associated with the fact that neither the function $G^{(2)}$ nor the probability (\ref{distsu}) nor the state $BS \vert z_E, 0 \rangle$ can be factorized.

As we have seen, the nonlinear coherent states $\vert z_E \rangle_N$ are not full coherent although their $P$-representation is a delta function, so that they cannot be considered `classical' in the sense established by Glauber \cite{Gla07}. Besides, the impossibility of factorizing either the second order correlation function $G^{(2)}$, or the probability of detecting $n$ and $m$ photons in the horizontal and vertical output channels, or the output state $BS\vert z_E, n \rangle$ associated with a 50:50 beam splitter, means that the fields represented by $\vert z_E \rangle_N$ are nonclassical.

%---------------------------------------> Section
\section{Concluding remarks}
\label{conclu}

We have shown that the nonlinear coherent states associated to a series of generalized oscillator algebras can be written in the same mathematical form. If such algebras are polynomial the related coherent states satisfy a closure relation that is uniquely expressed in terms of the Meijer $G$-function. We have also shown that, although the $P$-representation of these states is as singular as the delta function, they have no classical analog. The latter is due to the fact that such states are not full coherent in the sense established by Glauber in his quantum theory of optical coherence \cite{Gla07}. Then, a field represented by any nonlinear coherent state which is $P$-represented by a delta function will have classical analog whenever such state is full coherent.

As a byproduct of this work, we have shown that the criterion of separability introduced in \cite{Kim02} for distinguishing nonclassicality of states can be refined by considering also the separability of either the second order correlation function $G^{(2)}$, or the probability of detecting $n$ and $m$ photons in the horizontal and vertical output ports of a 50:50 beam splitter.

%---------------------------------------> Section
\appendix
\section{The $E$-exponential function}
\label{ApA}

\renewcommand{\thesection}{A-\arabic{section}}
% redefine the command that creates the equation no.
\setcounter{section}{0}  % reset counter 

\renewcommand{\theequation}{A-\arabic{equation}}
% redefine the command that creates the equation no.
\setcounter{equation}{0}  % reset counter 

The $E$-function introduced in Sec.~\ref{S_nonlinear},
\be
e^x_E = \sum_{n=0}^{\infty} \frac{x^n}{E(n)!}, \quad E(n)!= E(1) E(2) \cdots E(n), \quad E(0)! \equiv 1, 
\label{A1}
\ee
acquires a simple form if $E(n)$ is the polynomial of degree $\ell$ defined in (\ref{epol1})-(\ref{epol2}). Explicitly,
\be
e^x_E= {}_1F_{\ell}  \left( 1; 1+\delta_j; \frac{x}{\gamma_{\ell}} \right), \quad \delta_j = \frac{\beta_j}{\alpha_j}, \quad \gamma_{\ell} = \alpha_1 \cdots \alpha_{\ell}, \quad j =1, \ldots, \ell,
\label{A2}
\ee
where
\be
\begin{aligned}
  {}_pF_q (a_1, \ldots, a_p, b_1, \ldots, b_q;z)  \equiv & \,\,  {}_pF_q (a_j, b_j; z) \\[1ex]
= &    \frac{ \Gamma(b_1) \cdots \Gamma(b_q) }{ \Gamma(a_1) \cdots \Gamma(a_p) } \sum_{n=0}^{\infty} \frac{ \Gamma(a_1 + n) \cdots \Gamma(a_p +n) }{ \Gamma(b_1 + n) \cdots \Gamma(b_q +n)} \frac{z^n}{n!}
\end{aligned}
\nonumber
\ee
stands for the generalized hypergeometric function \cite{Olv10}. The following particular cases are of special interest:

\noindent
$\bullet$ For $\ell =1$ and $\beta \neq 0$ we have
\be
e^x_E= {}_1F_1 \left( 1, 1+\delta, \frac{x}{\alpha}  \right) = \delta e^{x/\alpha} \left( \frac{\alpha}{x} \right)^{\delta} \left[ \Gamma(\delta) - \Gamma \left (\delta, \frac{x}{\alpha}  \right) \right].
\label{A3}
\ee
In particular, for $\beta=0$ we get $\delta=0$, and
\be
e^x_E= \lim_{\delta \rightarrow 0} {}_1F_1 \left( 1, 1+\delta, \frac{x}{\alpha}  \right) = e^{x/\alpha}.
\label{A4}
\ee

\noindent
$\bullet$ For $\ell =2$ and either $\beta_1=0$ or $\beta_2=0$, one gets
\be
e^x_E= {}_1 F_2 (1; 1, 1+\delta; x) = {}_1 F_2 (1; 1+\delta, 1; x) = \Gamma(1+\delta) \left( \frac{\gamma_2}{x} \right)^{\delta/2} I_{\delta} \left( 2 \sqrt{ \frac{x}{\gamma_2}  } \right),
\label{A5}
\ee
with $I_{\nu}(z)$ the modified Bessel function of the first kind \cite{Olv10}.

\noindent
$\bullet$ For $\ell =3$ and $\alpha_1 = \alpha_2 = \alpha_3= \beta_1= \alpha$, with arbitrary $\beta_2$ and $\beta_3$, the $E$-exponential function (\ref{A1}) becomes
\be
e^x_E= {}_1 F_3 \left( 1;2, 1+ \delta_2, 1+\delta_3; \frac{x}{\gamma_3} \right) = \frac{\gamma_3 \delta_2 \delta_3}{x} \left[  {}_0 F_2 \left( \delta_2,\delta_3; \frac{x}{\gamma_3} \right) 
-1 \right].
\label{A6}
\ee

%---------------------------------------> Section
\section*{Acknowledgment}

We acknowledge the financial support from Instituto Polit\'ecnico Nacional, Mexico (Project SIP20170233), the Spanish MINECO (Project MTM2014-57129-C2-1-P) and Junta de Castilla y Le\'on (VA057U16). KZ and ZBG gratefully acknowledge the funding received through the CONACyT scholarships 45454 and 489856, respectively.

%---------------------------------------> Bibliography

\end{document}